
\input harvmac.tex
\font\Bigit=cmti10 scaled \magstep2
\noblackbox\parindent=0pt
\def\a{\alpha}\def\b{\beta}
\def\d{\delta}\def\D{\Delta}
\def\z{\zeta}\def\t{\theta}

\def\l{\lambda}
\def\O{\Omega}

\def\pa{\partial}\def\na{\nabla}
\def\phidot{{\dot\phi}}
\font\large=cmr10 scaled\magstep 1
 2
 3
 3
\rightline {UPR-522T}
\bigskip
\centerline{{\large STRINGY INSTANTONS}}
\vskip.3in
\centerline {RAM BRUSTEIN and BURT A. OVRUT}
\bigskip\baselineskip=10pt
\centerline{Department of Physics}
\centerline{University of Pennsylvania}
\centerline{Philadelphia, PA 19104}
\vskip.8in

\baselineskip=12pt plus 2pt minus 1 pt
 {\it \centerline{ABSTRACT}
\bigskip
\noindent
A  canonical Lorentz invariant field theory
extension of  collective field theory of $d=1$ matrix models is presented.
 We show that the low density, discrete, sector of collective field theory
includes single  eigenvalue Euclidean instantons which tunnel between
different vacua of the extended theory. These ``stringy" instantons induce
non-perturbative  effective operators of strength  $e^{-{1/g}}$.
The relationship of the world sheet description of string theory and
Liouville theory to the effective space-time theory is explained.}
\bigskip\bigskip

\footnote{}{{\sevenrm email: ramy@mohlsun.physics.upenn.edu}}
 \footnote{}{{\sevenrm email: ovrut@penndrls.upenn.edu}}

\footline={\hss\tenrm\folio\hss}
\baselineskip=16pt plus 2pt minus 2pt

Non-perturbative aspects of string theory are an essential piece
of information needed to confront string theory predictions and the real
observable world.

Matrix models, and especially $d=1$ matrix models
\ref\grossa{ D. J. Gross, N. Miljkovic,  Phys.Lett.B238 (1990) 217;
 P. Ginsparg and J. Zinn-Justin, Phys. Lett. B240 (1990) 333;
 E. Brezin, V. Kazakov and Al. B. Zamolodchikov, Nucl. Phys. B338 (1990) 673.},
 offer  a unique opportunity to obtain some  insight into non-perturbative
string theory.
The $d=1$ matrix model is the most complicated matrix model
which can be explicitly solved.
On the other hand, it describes the simplest space-time dynamics which
is still interesting. In the double scaling limit, the $d=1$ matrix model
 describes strings propagating in one time dimension
and one spatial dimension. An equivalent description is given in terms
of a bosonic collective field theory in $1+1$ dimensions of one massless field
\ref\das{S.R. Das and A. Jevicki, Mod. Phys. Lett. A5 (1990) 1639.},
\ref\joeb{ J. Polchinski,  Nucl.Phys. B346 (1990) 253.}. A notable feature
of collective field theory is that  the kinetic energy is not in canonical form
and it is not Lorentz invariant.
The $d=1$ matrix models, or the equivalent field theories
have the power to describe non-perturbative phenomena in the associated
$1+1$ string theories. By studying the generic features of non-perturbative
behaviour in $1+1$ dimensional string theories, as we do in this letter,
one may learn about more realistic 4-dimensional string theories.

Large  order growth of perturbative
amplitudes is  a common feature of matrix models
 and more complicated string theories
\ref\shenker{S. Shenker, Cargese Workshop on Random
Surfaces, Quantum Gravity and Strings, Cargese, France (1990).}.
For a review of large order behaviour of matrix model amplitudes
see ref.\ref\zinn{ P. Ginsparg and
J. Zinn-Justin, Phys. Lett. B255 (1991) 189.}.
All matrix models, as well as the critical bosonic string theory in 26
dimensions, exhibit a strange phenomenon. The magnitude of $G$'th order
amplitudes in perturbation theory  grow like $(2G)!$.

In the same way  as $G!$ behaviour corresponds to well understood
$e^{-{1\over g^2}}$ non-perturbative effects in quantum field theory,
in matrix models the large order  $(2G)!$ behaviour
would correspond to peculiar
non-perturbative effects of strength $e^{-{1\over g}}$.
How do these peculiar effects arise?
It turns out that in matrix models, there is a new type of instanton, involving
a single eigenvalue, that is responsible for these effects.
For a discussion of one eigenvalue instantons see \shenker,\zinn.
They were also discussed in ref. \ref\lecht{O. Lechtenfeld,
Int. J. Mod.Phys. A7 (1992).} and in the
context of supersymmetric matrix models in refs.
\ref\dabh{A. Dabholkar, Nucl.Phys. B368 (1992) 293.} and
\ref\hans{ J. D. Cohn and  H. Dykstra, Mod.Phys.Lett. A7 (1992) 1163.}.
We show that in matrix models, the associated effective field theory action
does not obey the usual field theoretic  scaling argument,
$S(\phi,g) \ne {1\over g^2} \widetilde S(\widetilde\phi)$. Instead,
one finds that $g$ cannot be completely scaled out of $\widetilde S$
due to ``scale breaking terms". That is
$S(\phi,g) = {1\over g^2} \widetilde S(\widetilde\phi,g)$. It follows
that a non-trivial solution can be a function of $g$. Furthermore, if
for such a solution $\widetilde S\sim g$, then $S\sim {1\over g}$.
This is exactly what happens for one eigenvalue instantons.

In this letter we  discuss the high and low density
limits of collective field theory. We show that, in Euclidean space,
there is a single eigenvalue instanton solution of the equations of motion
which has action ${\pi\over g}$. We then extend the collective field theory,
which is a non-canonical, non-Lorentz invariant theory of a single
field $\phi$,
to a canonical, Lorentz invariant effective field theory of two
fields $\z$ and $D$. This theory exhibits the scale breaking terms
responsible for the unusual action, $S\sim {1\over g}$, of the instantons.
We show that the vacuum solutions of the
effective field theory includes wormhole-like configurations.
These configurations look like two identical Liouville vacua linked by
a single eigenvalue instanton. The normalized action of these
configurations is $S={\pi\over g}$. We then show that these configurations
induce calculable operators in the $\z$, $D$ effective theory with strength
$e^{-{\pi\over g}}$, as conjectured by Shenker. We also review and explain
the relationship of these double scaled matrix models to $1+1$ dimensional
string theories, and argue that these string theories must include the
new vacuum configurations, single eigenvalue instantons, and induced
operators of strength $e^{-{\pi\over g}}$. An extended and detailed discussion
of all the issues in this letter and more can be found in
\ref\ramburt{R. Brustein and B. Ovrut, Penn preprint, UPR-523T (1992).}.

A useful step on the road from  the matrix model to string theory is
 collective field theory \das. Relevant references are
\ref\sakicki{B. Sakita, Quantum Theory of Many-Variable Systems and
Fields, World Scientific, Singapore 1985.}--\nref\cohn{J. D. Cohn
and S.P. De Alwis, Nucl.Phys.B368 (1992) 79.}\ref\karb{D. Karabali
and B. Sakita,
 Int. J. Mod. Phys. A6 (1991) 5079.}.
The idea is to  start from the matrix model and, by
performing a series of changes of variables, arrive at a field theory
representation of the matrix model. The collective field  is
$\phi(x,t)=\sum\limits_i\d(x-\l_i(t))$  where $\l_i,\ i=1,...,N$
are the eigenvalues of the matrix. The partition function in terms of the
collective field is
\eqn\partcoll{ Z_N(g_n) =
  \int [d\phi]\ {\rm det}\O^{-\half}
\hbox{\it e}^{i  N^2\int dt dx \left\{
\half{\int\limits^x\phidot \int\limits^x\phidot \over \phi}
-{\pi^2\over 6} \phi^3- (V(x)-\mu_F)\phi \right\}      }   }
Here  contact terms are omitted.
Note the appearance of the factor ${\rm det}\O^{-\half}$ which comes from
doing the  Gaussian integral over the conjugate momentum $\Pi_\phi$.
The eigenvalues $\l_i$ are independent variables,
However, for finite $N$ the collective field
$\phi(x,t)$ is highly constrained. It does  not correspond to an infinite
number of degrees of freedom. Hence, the $\phi$  equation of motion is
 not obtained by varying the action with respect to $\phi$.
The correct procedure is the following.
Start with the equations of motion for the eigenvalues,
then by using the relation between $\phi$ and $\l_i$ convert them
to an equation for $\phi$. The result is
\eqn\motphi{ {\pa\over\pa y} {\d S_{\rm eff}[ {\dot\phi},\phi]\over
\d\phi(y,t) }{}_{|_{y=\l_j(t)}}=0 \ \ \ \ j=1,...,N }
Note that there are $N$ equations of motion since $y$ must be evaluated
at $\l_j$ for all $j=1,...,N$. Furthermore, note that $\phi's$ satisfying the
naive $\phi$ equation of motion
${\d S_{\rm eff}[ {\dot\phi},\phi]\over
\d\phi(y,t) } =0$ also satisfy  Eq.\motphi. However there are solutions
of Eq.\motphi\ which do not satisfy the naive equation of motion.

Consider now the limit $N\rightarrow\infty$.
For our purposes it is general enough to consider a potential
$V(x)=V(0)-\half x^2$.
Inserting this expression into the Lagrangian of Eq.\partcoll,
the combination $V(0)-\mu_F$ appears.
We denote $V(0)-\mu_F$ by $\half\mu$ and  assume that $\mu>0$.
The double scaling limit is defined by specifying the $N$ dependence of $\mu$,
$N\mu={1\over g}$,
so that $g$ remains finite as $N\rightarrow \infty$.
The classical equations of motion derived using Eq.\motphi\
are
\eqn\eqmotcoll{ {\pa\over \pa x} \left(
\int\limits^x dy \pa_t{\int\limits^y\phidot\over\phi}
-{1\over 2}{\int\limits^x\phidot \int\limits^x\phidot \over\phi^2}
 -{\pi^2\over 2} \phi^2 -\half ({1\over g}-x^2)\right){}_{|x=\l_i(t)} =0 }
where the index $i$ now runs over $i=1,...,N\rightarrow\infty$.

The high density (HD) limit of collective field theory is defined as follows.
Consider  a region of  $x$, denoted $I$,  of length
$l(I)$. The number of
eigenvalues in this region is $N(I)$.
Then take the  double scaling limit in such a way that
${N(I)\over l(I)}\rightarrow\infty$.
It is clear that in this region of $x$ there are an infinite number of
eigenvalues. The collective field $\phi$ now has an infinite number of degrees
of freedom  and the classical equations of motion  in this region are
 Eq.\eqmotcoll\ without the restriction $x=\l_i(t)$.
The static solution of these equations is
$\phi_0= {1\over\pi}\sqrt{x^2-{1\over g}}$,
where $|x|\ge \sqrt{{1\over g}}$. Note that $\phi_0$ actually
 makes the term inside the parenthesis in  Eq.\eqmotcoll\ vanish.
That is, $\phi_0$ is a solution of the conventional field theory
equations  of motion. The coupling parameter of collective field theory is
 ${1\over \pi^{3\over 2} \phi_0^2(x)}$. However, ${1\over\phi_0^2(x)}$ becomes
large exactly where the classical solution $\phi_0$ becomes small, which is
where the HD expansion breaks down. The field theory becomes strongly
coupled or in other words the semi-classical expansion breaks down.
What happens if this fact is ignored? Perturbation series  takes its revenge
by growing too fast. In fact, the $G$'th order in perturbation
expansion grows like $(2G)!$. We take the HD region to be the region in x
space where the static solution $\phi_0$ is defined; that is
$|x|\ge \sqrt{{1\over g}}$.

The low density  (LD) limit of collective field theory is defined as follows.
Consider  a region of  $x$, denoted $J$,  of length
$l(J)$. The number of eigenvalues in this region is $N(J)$.
Then take the  double scaling limit in such a way that
${N(J)\over l(J)}\rightarrow finite$.
It is clear that in this region of $x$ there are a finite number of
eigenvalues. The collective field $\phi$ now has a finite number of degrees
of freedom  in that region
 and the classical equations of motion are  Eq.\eqmotcoll\
where the index $i$ now runs over $i=1,...,N(J)$.

In this letter, we are particularly concerned with LD regions containing a
single eigenvalue, $\l^*$, in Euclidean space. Going to Euclidean time, $\t$,
and using the relation $\phi(x,\t)=\d(x-\l^*(\t))$
the equation of motion  (including the contact terms) simply becomes
${d^2\l^*\over d\t^2} =  -\l^*(\t)$.
The general solution of this equation is given by
$\l^*(\t)=A^* cos \t+ B^* sin \t$
where $A^*$ and $B^*$ are real constants.
The Euclidean  LD region is taken to be
$-\sqrt{{1\over g}}\le x\le \sqrt{{1\over g}}$.

As an example consider the collective field  corresponding to a
single eigenvalue $\l^*$ with the boundary conditions
$\l^*(\t_0)=\sqrt{{1\over g} }$ and ${\dot\l}^*(\t_0)=0$. It follows  that
$\l^*(\t) =\sqrt{{1\over g}} cos(\t-\t_0)$.
The corresponding collective field configuration is
\eqn\oneinst{
\phi_{inst}(x,\t)=\d\left(x-{1\over\sqrt{g}} cos(\t-\t_0)\right) }
This is an instanton that corresponds to tunneling of one eigenvalue
 across the barrier from
$x=\sqrt{{1\over g}}$ at $\t=\t_0$ to
$ x= -\sqrt{{1\over g}}$ at $\t=\t_0+\pi$. Note that the classical
solution in Eq.\oneinst\ is not a solution of the Euclidean continuation
of the unconstrained  field theory equations of motion.

The action of the instanton $\phi_{inst}$
 can be computed from the Euclidean continuation of \partcoll. The result is
\eqn\actoneinst{S_{\rm eff}[\phi_{inst},\phidot_{inst}]=\int_0^{\pi} d\t
\int\limits_{+\sqrt{{1\over g} }}^{-\sqrt{{1\over g} } }dx
\d\left(x- {1\over\sqrt{ g}}
cos(\t)\right)\left\{ {1\over 2 g}\sin^2(\t) -{1\over 2 g}\cos^2(\t) +
{1\over g} \right\}= {\pi\over g} }
in agreement with the large order behavior of the perturbation series in $g$.

We begin the discussion of the effective field theory in the HD region
$x\ge\sqrt{{1\over g}}$.
The first step is to shift the double scaled collective field $\phi$ by the
classical static solution $\phi_0$,
$\phi=\phi_0+{1\over\sqrt{\pi}}\pa_x\z$. To obtain a canonical kinetic term
for the field $\z$,  change coordinates to the Liouville coordinate defined by
$\tau-\tau_0^I
={1\over\pi}\int\limits^x_{x_0} {dy\over\phi_0}
=\ln\left[x+\sqrt{x^2-{1\over g} }\ \right]
-\ln\left[x_0+\sqrt{x_0^2-{1\over  g}}\ \right]$. For simplicity, we take
$x_0=\sqrt{{1\over g} }$ and $\tau_0^I=\ln\sqrt{ {1\over g}} $. In this case
$x= \sqrt{{1\over g}}\cosh(\tau-\ln\sqrt{{1\over g}})$.
and the range of  the two coordinates is
$ \sqrt{{1\over {g}}}\le  \ x\ \le \infty $ and
$\ln[\sqrt{{1\over {g}}}\ ] \le \  \tau\ \le \infty$.
The static solution in the new  coordinate becomes
$\phi_0(\tau)= {1\over \pi \sqrt{g}}\sinh (\tau-\ln\sqrt{{1\over g}})$.
Rewritten in terms of the Liouville coordinate the classical
Lagrangian is $L=L_\z(\z,\phi_0)+L_0(\phi_0)$ where $L_\z$ is given by
\eqn\lzeta{ L_\z  =\int d\tau\left\{
 \half{{\dot\z}^2 \over 1+{1\over \pi^{3\over 2}\phi_0^2(\tau)} \pa_\tau\z}
-\half (\pa_\tau\z)^2
-{1\over 6 }{1\over \pi^{3\over 2}\phi_0^2(\tau) } (\pa_\tau\z)^3\right\} }
and the pure $\phi_0$ Lagrangian, $L_0$, turns into
$L_0=\int d\tau {\pi^3\over 3}\phi_0^4$.
{}From the cubic interaction term in $L_\z$, it follows that
the coupling parameter of collective field theory is
${ 1\over \pi^{3\over 2}\phi_0^2(\tau)}=
4\sqrt{\pi} {e^{-2\tau}\over\left(1-{1\over {g}} e^{-2\tau} \right)^2}$.
The coupling parameter vanishes as $\tau\rightarrow\infty$ and explodes at
$\tau=\ln[\sqrt{{1\over {g}}}\ ]$.

The low density version of collective field theory is obtained from
Eq.\partcoll\
by setting $\phi=\pa_x\z$. The resulting Lagrangian is
\eqn\lzetaII{L_\z=\int dx \left\{ { {\dot\z}^2\over\pa_x\z }
-{\pi^2\over 6} (\pa_x\z)^3 -\half({1\over g}-x^2)\pa_x\z\right\} }

We now describe a field theory that reduces to  the collective field
theory  in the region $\tau\ge\ln\sqrt{{1\over g}}$
 when the various fields obtain their expectation values.
The idea was discussed  for  $\mu=0$  in
\ref\ramyshanta{R. Brustein and S. De Alwis, Phys. Lett. B272 (1992) 285.}.
We  note that the $\z$ theory is not Lorentz invariant. Our interpretation is
that this is really a Lorentz invariant field theory of two fields,
$\z$ and $D$,
expanded around  the vacuum expectation values of the two fields.
The new field $D$ has a vacuum expectation value that breaks Lorentz
invariance, and that is the reason that the $\z$ theory alone is not
Lorentz invariant.

To find out the background independent field theory we have to identify
the expectation value of the $\z$ and $D$  fields first.
Motivated by the comparison between collective field theory and the
Polyakov description of the related string theory
we postulate that
\eqn\ohsol{<G_{\mu\nu}> =\eta_{\mu\nu},\ \ \ <D>  = - 2\tau,\ \ \
<\zeta> ={1\over g} }
Here we added the expectation value of the metric as well. Note that
the field  $D$ has the non-translation invariant vacuum expectation value.

We list the background independent form of the different quantities.
$\pi^{3\over 2}\phi_0^2(\tau)\rightarrow
{1 \over 4\sqrt{\pi} } e^{-D}\left(1-{1\over {g}} e^{D} \right)^2$,
$\pa_\tau \z \rightarrow \half \na D\cdot\na\z$,
${\dot\z}^2-(\pa_\tau \z)^2\rightarrow\na\z\cdot\na\z$.
The Lorentz non-invariant quantities on the left are obtained from the Lorentz
invariant quantities on the right by letting $D=<D>$ and $\z=<\z>+\z'$.
Also, let $\int dt d\tau=\int d^2x$.
We can now write the $D,\z$ action using the previous dictionary of
expressions. The result is
\eqn\monster{\eqalign{
 {\cal S}=  \int d^2x \Biggl\{&
\half{\na \z\cdot\na \z\over 1+{ 2\sqrt{\pi} }
{ e^{D}\over\left(1- {1\over {g}} e^{D} \right)^2 }\na \z\cdot\na D}
- { \sqrt{\pi}\over 4 }{ e^{D}\over\left(1- {1\over {g}} e^{D} \right)^2 }
{(\na \z\cdot\na D)^3\over 1+{ 2 \sqrt{\pi} }
{ e^{D}\over\left(1-{1\over {g}} e^{D} \right)^2 }\na \z\cdot\na D}
\cr  -&  { \sqrt{\pi}\over 12}
{e^{D}\over\left(1-{1\over {g}} e^{D} \right)^2 }
(\na \z\cdot\na D)^3 -
{1\over  384 \pi} e^{-2 D}
 \left[1-{1\over {g}} e^{D} \right]^4
\left[ {(\na D)^2-4} \right]
\Biggr\} \cr }  }
Note that $\z$ has only derivative couplings and therefore the action
\monster\ is invariant under a shift symmetry $\z\rightarrow\z+const.$
Since space-time is flat in this case  we ignore  curvature terms in the
action. Higher derivative terms like $(\na D\cdot\na D)^2$ cannot be ruled
out at this point. In what follows we treat the  above Lagrangian  as
if it were the exact Lagrangian. It is straight forward to verify that indeed
the field configurations in Eq.\ohsol\ is an exact solution of the
 equations of motion.

It is important to note that all interaction terms in
\monster\ are proportional to
$ \hbox{\Bigit g} (D)=4\sqrt{\pi}
 { e^{D}\over\left(1- {1\over {g}} e^{D} \right)^2 }$
and, therefore ${\hbox{\Bigit g}} (D)$ is the effective  coupling parameter
of the theory. The general solution of the equations of motion is given by
\eqn\sssolmina{<G_{\mu\nu}>  =\eta_{\mu\nu},\ \ \
<D>  =   a (t-{\bar t}) + b (\tau -{\bar\tau}),\ \ \
<\z>  ={1\over g}+c }
where $a,b,c,{\bar t}$ and ${\bar\tau}$ are real parameters, $b^2-a^2=4$ and
$c,{\bar t},{\bar\tau}$ are arbitrary.

There is an interesting  vacuum structure
which is the combination of  two solutions. That is, take $<\z>={1\over g}$
 and $<D>=-2\tau$
for $\tau\ge\tau_0^I=\ln\sqrt{{1\over g}}$, henceforth called region $I$,
and $<D>=-2(\tau-2 \ln\sqrt{{1\over g}}+\pi)$
for $\tau\le\tau_0^{III}=\ln\sqrt{{1\over g}}-\pi$, henceforth called region
$III$. The spatial interval
$\ln\sqrt{{1\over g}}-\pi\le \tau\le \ln\sqrt{{1\over g}}$
is called region $II$. We plot these regions in Figure 1.

The physical interpretation of this vacuum state is the following.
In regions $I$
and $III$, away from the boundary points, the effective coupling parameter
is small and physics is described by the effective field theory \monster.
As $\tau$ approaches $\ln\sqrt{{1\over g}}$ from the right and as $\tau$
approaches $\ln\sqrt{{1\over g}}-\pi$ from the left, the coupling parameter
blows up and a region of strong coupling is encountered. Region $II$ is
terra incognita. Perhaps new, previously unknown, dynamics applies there.

 If all we knew was the effective
field theory, then we would have no interpretation of physics in region $II$.
However, comparing the vacuum of Figure 1 to the matrix model
solution, we know exactly how to describe physics in region II. Physics
in that region is  described by  the low density collective field theory.

The action for Euclidean space-time effective field theory is easily
obtained from Eq.\monster\ by analytic continuation of the time variable
$t$ to Euclidean time $\t$. We do not write it down explicitly.
The effective  coupling parameter and the static solutions of the
Euclidean equation of motion  are still given by the Euclidean continuation
of the same expressions as before.
It follows that in Euclidean space, the vacuum structure Figure 1 is  valid.
Region $II$ is now described by the analytic continuation of the low density
collective field theory to Euclidean space. There is now a non-trivial
excitation of one eigenvalue in region $II$ that connects the vacua of
region $I$
and region $III$. This single
eigenvalue excitation in Euclidean space was constructed previously, and
presented in terms of the collective field theory in Eq.\oneinst.
Rewriting this solution in  the $\tau$ coordinate, we find
\eqn\instburt{\phi_{inst}(\tau,\t)=
{1\over\sin(\t-\t_0) }
\d\left (\tau-[\ln\sqrt{{1\over g}}-(\t-\t_0)]\right) }
This is an instanton which corresponds to the tunneling of a single
eigenvalue across the barrier from $\tau=\ln\sqrt{ {1\over g}}$ at
$\t=\t_0$ to $\tau=\ln\sqrt{ {1\over g}}-\pi$ at $\t=\t_0+\pi$. Note that
the velocity of the eigenvalue
at either side of the barrier vanishes. Therefore,
the Euclidean conjugate momentum of the
instanton in region $II$, matches continuously at the boundaries with the
vanishing  conjugate momentum of the static  vacua $\phi_0$
in regions $I$ and $III$.
We represent this tunneling process in Figure 1.
\bigskip
\ \vskip2.5in
%
%
%
\centerline{Figure 1. Instanton Connecting regions $I$ and $III$.}
\bigbreak
In terms of $\z$ the instanton is given by
$\z_{inst}(\tau,\t)=
\Theta \left(\tau-[\ln{1\over\sqrt{g}}-(\t-\t_0)]\right)$.
Therefore, the instanton field configuration is simply
 a kink moving in Euclidean time.
The position of the kink is where the argument of the  $\Theta$ function
vanishes.

We want to stress that this configuration is not a solution of the
$D$, $\z$ effective field theory.  We have not included explicitly the
terms in the field theory that correspond to self energy subtraction.
However once they are taken into account, the instanton configuration
has a finite action ${\pi\over g}$, as shown in Eq.\actoneinst.
Finally, we note that the initial tunneling time of the instanton, $\t_0$
is arbitrary.

We  integrate over the instantons and represent
their effects as effective terms in the $D$,$\z$ theory. Since $\z$ is
the light field we restrict our attention to $\z$ operators.
The effective operators are especially important. They provide the
bridge between the low density, discrete, sector of the theory and the
continuous sector. A more detailed analysis is given in  \ramburt.

The  instanton has three parameters, $\bar\tau$, $\t_0$ and $\alpha$.
The parameters $\bar\tau$ and $\t_0$ were  defined in Eqs.\sssolmina\
and \instburt. The parameter $\a$ is  related to the  parameter
$a$ in the Euclidean space continuation of Eq.\sssolmina,   $a=2\sin\a$.
Changing $\a$ results in the rotation of the vacuum solution in
$\tau-\t$ space.
There are three zero modes corresponding to  the three broken generators
of the Euclidean group associated with $\bar\tau$, $\t_0$, $\a$.
These have to be integrated and produce a volume
factor $Vol\propto \int d\bar\tau d\t_0  d\a$.

The dilute gas summation
over instantons induces  effective terms in the $D$,$\z$ Lagrangian.
The most general  action   induced by instantons is
$\Delta S=\int d\tau d\t  \{\sum\limits_n C_n O_n(\tau,\t)\}$,
where $O_n$ are local operators built from  $D$ and $\z$ and
their derivatives.
The coefficients $C_n$ can be computed by expanding the action around the
instanton background.
All the coefficients $C_n$ are proportional to the universal factor of the
exponent of the instanton action  and the remaining
factor depends on the  particular operator that is considered.
 Since the ``size" of the instanton is $\sqrt{{g}}$,
the dimension of the operator determines the $ g$ dependence of $C_n$. That is
\eqn\ssss{C_n= \hbox{\it \~C}_n g^{d(n)} e^{-{\pi\over g} }}
where $d(n)=[{\rm dimension}(O_n)]^{\half}-1$
and $\hbox{\it \~C}_n$ is a numerical coefficient.  The coefficient
$\hbox{\it \~C}_n$ is not expected to be particularly
large or particularly small.

We are interested in large ${1\over g}$ that corresponds to small $g$.
In that case the dominant and most interesting operator  is the unit operator.
All other operators are suppressed by powers of $ g$.
The coefficient  of the unit operator is given by
$C_0= \hbox{\it \~C}_0 {\scriptstyle {1\over g}} e^{-{\pi\over g} }$.
This result  was obtained in the background of a constant
field $<\z>={1\over g}$. Lorentz
invariance then dictates that at least for slowly varying fields the effective
operator depends on the full field $\z$ and not just its constant
mode ${1\over g}$.
Therefore the final result for the  induced operator is
\eqn\effop{ \D{\cal L}_0=\hbox{\it \~C}_0\z e^{-\pi\z}}
This operator breaks the  $\z$ shift symmetry. It induces a  runaway
non-perturbative potential for the field $\z$.

Let us now discuss the relation of the effective field theory \monster\
to string theory.
The class  of $1+1$ dimensional string theories that  we are  interested  in
is  described  by the  following two dimensional $\sigma$-model
\ref\eli{S. Elitzur, A. Forge and E. Rabinovici,
Nucl. Phys. B359 (1991) 581\semi
            A. Tseytlin, Phys. Lett. B264 (1991) 311.},
\eqn\sigac{I=
{1\over 4\pi}\int d^2z\sqrt{\hat g}
\left\{\hat g^{\a\b}G_{\mu\nu}\pa_{\a}X^{\mu}
\pa_{\b}X^{\nu} +\hat R D(X)+2 T(X)\right\}}
where $\hat g_{\a\b}$ is the fixed world sheet metric  with Euclidean
signature and  $\hat R$ is the corresponding Ricci scalar.
The sigma model  field $X_\mu$  stands for  two  scalar
fields, $X_0 (z)$, and $X_1(z)$.
The   field $G_{\mu\nu}(X)$ is the target space metric, assumed here to have
Euclidean signature, $D(X)$ is the dilaton,  and $T(X)$ is the tachyon.

Consistent  string backgrounds  are described by conformal field  theories.
The conditions for  conformal invariance are determined in general  by
the equations $\beta=0$. The lowest order equations for the theory described
by Eq.\sigac\  are
\eqn\bz{R_{\mu\nu}+2\nabla_\mu\nabla_\nu D  =  0 ,\ \ \
                  -\half \nabla^2 D+ (\nabla D)^2 + 4  =  0,\ \ \
                  -\nabla^2 T+2 \nabla D\cdot\nabla T-4 T  =  0 }
We can compare Eqs.\bz\ to the lowest order (in $\hbox{\Bigit g}(D)$)
equations  of motion
derived from \monster. From this comparison we deduce that, to this order,
the field $D$ appearing in \monster\ is the same as the dilaton $D$
in \sigac\ and that the field $\z$ in \monster\ is related to the tachyon,
 $\z\propto   T  e^{-D}$.
This result that we obtained is rather remarkable.
It says that the equations of motion
of the background fields for the class of string theories specified by
the action \sigac\ are, to lowest order, identical to the equations
of motion of the effective field theory extension of matrix models. Can we
extend this identification beyond lowest order? To do this, let us consider
the solutions of the equations of motion \bz. The flat space solutions are
$G_{\mu\nu}  =  \d_{\mu\nu}$, $D  = a(X_0-\bar X_0)+ b(X_1-\bar X_1)$,
$T    = m e^D$,
where $a^2+b^2=4$. Of particular interest is the static  solution
$G_{\mu\nu}  =\ \d_{\mu\nu}$, $D\   =-2 X_1$,  $T\ = m e^{-2 X_1}$.
By substituting this static  solution  into the sigma model
\sigac\ and writing $X_1=\varphi$, the following  world sheet
conformal field  theory is  obtained
\eqn\lcft{ I=
{1\over 4\pi}\int d^2z\sqrt{\hat g}\left\{\hat g^{\a\b} \pa_{\a}X_0\pa_{\b}X_0
+ \hat g^{\a\b} \pa_{\a}\varphi\pa_{\b}\varphi-2 \hat R \varphi
+2 m e^{-2\varphi} \right\}}
which can be identified as the Liouville conformal field theory with $c_m=1$
matter. To obtain a Minkowski signature string theory one has to analytically
continue $X_0\rightarrow i\hbox{\Bigit t}\ \!(z)$, which then becomes
the time variable of target space. The field $\varphi$ corresponds to
the spatial dimension of target space.

Liouville  theory \lcft\  was  compared  to matrix  models and to  collective
field theory. Specifically, the complete theory, beyond the lowest order
was studied. The  conclusion is that they
describe of the same theory. Evidence to this effect was obtained
on many  levels e.g., see
\ref\igor{D. J. Gross and I.R. Klebanov,
Nucl.Phys. B352 (1991) 671.}--\nref\grig{D. J. Gross and I. R.  Klebanov,
Nucl. Phys. B359 (1991) 3.}\nref\huld{ G. Moore,
Nucl.Phys. B368 (1992) 557.}\nref\antalb{K. Demeterfi, A. Jevicki and
 J. P. Rodrigues, Nucl.Phys. B362 (1991) 173 and
Nucl.Phys. B365 (1991) 499.}\nref\antalc{K. Demeterfi, A. Jevicki and
J. P. Rodrigues, Mod. Phys. Lett. A6 (1991) 3199.}\ref\fraku{P. Di Francesco,
D. Kutasov, Nucl. Phys. B375 (1992), 119.}.
For a review and more comprehensive list of references see ref.\antalc.
In particular, the  relation between $\z$  and $T$ as well
as the linear relation between $\varphi$ and $\tau$ have been well
documented.

More important, from our point of view, is that, by computing scattering
amplitudes of fluctuations around the Liouville vacuum,
one can determine the equations of the original string backgrounds $D$ and $T$
beyond the lowest order. When these are compared to the full equations of
motion derived from Eq.\monster, one finds that they are identical, as long as
the two $D$ fields are identified, $\z=T e^{-D}$ holds and the parameter
 $m$ in  is chosen to be $m={1\over g}$.
We conclude, therefore, that
the string theories associated with the world
sheet action \sigac\  have the same equations of motion and
effective action for their background fields as does the effective
field theory for the matrix model given in Eq.\monster.
 Furthermore, even the low density region of the matrix model discussed
earlier is  expected to to describe physical
aspects of these string theories, such as their non-perturbative behaviour.
It follows that the discussion of vacua, single eigenvalue instantons and
induced operators given in the previous section is, in fact, applicable
to the string theories associated with \sigac. The instanton in Eq.\instburt\
is therefore  a ``stringy" instanton.

It is tempting to conjecture that stringy instantons similar to our
stringy instantons appear in 4 dimensional superstring theories and that they
induce non-perturbative operators of the type discussed previously.
In that case these operators are expected to be proportional to
the universal factor $e^{-\sqrt{S}}$,  where $S$ is a complex field
that naturally appears in the effective low energy supergravity  field
theory obtained from superstring theory. The dilaton is related to
the real part of $S$, $<Re S>\sim{1\over g^2}$.
Note that the non-perturbative effects considered previously in the
literature induced operators that were proportional to the  universal factor
$e^{-S}$.
Since the coupling parameter $g$ is expected to be small, the difference
between these two universal factors is quite big.
This may have important phenomenological consequences.

{ACKNOWLEDGEMENT}:
It is a pleasure to thank Lee Brekke, Joanne Cohn, Rick Davis,
Antal Jevicki and especially  Shanta De Alwis for useful discussions.
This work was supported in part by the Department of Energy under
contract No. DOE-AC02-76-ERO-3071.

\listrefs
\bye